\documentclass{aa-package/aa}
\usepackage{graphicx}
\usepackage{txfonts}
\begin{document}
  \title{Radiative diffusion in stellar atmospheres: diffusion velocities }

  \author{G.~Alecian\inst{1} \and M.J. Stift\inst{2}\inst{,1} }

  \offprints{G. Alecian}

  \institute{LUTH (Observatoire de Paris - CNRS), 
             Observatoire de Meudon, F-92195 Meudon Cedex, France\\
             \email{georges.alecian@obspm.fr}
             \and
             Institut f{\"u}r Astronomie (IfA), Universit{\"a}t Wien,
             T{\"u}rkenschanzstrasse 17, A-1180 Wien\\
             \email{stift@astro.univie.ac.at}
             }

   \date{Received ...; accepted ...}

%  (context), aims, methods, results (conclusions)
   \abstract
     {}
     {The present paper addresses some of the problems in the
     buildup of element stratification in stellar magnetic 
     atmospheres due to microscopic diffusion, in particular
     the redistribution of momentum among the various ionisation
     stages of a given element and the calculation of diffusion 
     velocities in the presence of inclined magnetic fields.}
     % methods
     {We have considerably modified and extended our CARAT code to provide
     radiative accelerations, not only from bound-bound but also from
     bound-free transitions. In addition, our code now computes ionisation
     and recombination rates, both radiative and collisional.
     These rates are used in calculating the redistribution
     of momentum among the various ionisation stages of the chemical
     elements. A careful comparison shows that the two different
     theoretical approaches to redistribution that are presently
     available lead to 
     widely discrepant results for some chemical elements, especially 
     in the magnetic case. In the absence of a fully satisfactory 
     theory of redistribution, we propose to use the geometrical mean 
     of the radiative accelerations from both methods.}
     % results
     {Diffusion velocities have been calculated for 28 chemical
     elements in a $T_{\rm eff} = 12000\,K$, $\log g = 4.00$ stellar
     magnetic atmosphere with solar abundances. Velocities and 
     resulting element fluxes in magnetic fields are discussed; 
     rates of abundance changes are analysed for systematic trends with 
     field strength and field direction. Special consideration is 
     given to the Si case and our results are confronted in detail 
     with well-known results derived more than two decades ago.}
     {}

   \keywords{Diffusion -- 
   	Stars: abundances -- 
   	Stars : chemically peculiar --
   	Stars : magnetic fields
             }

\titlerunning{Diffusion velocities}
\maketitle

\section{Introduction}

Having been proposed more than 3 decades ago by Michaud (\cite{mic70}),
the diffusion theory still constitutes the only viable scenario
for the buildup of abundance inhomogeneities in chemically
peculiar (CP) early type main-sequence stars. Early calculations 
by Michaud et al. (\cite{mic76}) have shown that diffusion in the envelope 
of main-sequence stars can lead to overabundances of a 
number of chemical elements by up to 7\,dex, which would be sufficient 
to explain even the largest overabundances found in CP stars.

Magnetic fields can be expected to modify diffusion, and indeed Vauclair et al. 
(\cite{vhp79}) outlined a particular mechanism involving upward 
diffusion of Si\,{\sc i} that produces overabundances in magnetic
stars at places where the magnetic field is horizontal. Whether
vertical or horizontal magnetic fields could lead to the development of 
spots or rings with enhanced elemental abundances depends on the
element in question; silicon appears to concentrate in horizontal
field regions (Alecian \& Vauclair \cite{alv81}), whereas gallium is 
thought to accumulate near vertical field regions (Alecian \& Artru 
\cite{ala87}).

The detailed modelling of 53 Cam by Babel \& Michaud (\cite{bam91b}) has
shown that a simple diffusion model is not always sufficient and that
other processes should be considered, like inhomogeneous mass-loss.
Whenever one invokes diffusion of chemical elements as the mechanism
responsible for abundance anomalies in a star, this implies that a
stratification process has been at work somewhere in the star or is
still at work, depending on the stellar age and its fundamental
parameters. Observational evidence for such vertical abundance
stratification of chemical elements has accumulated over recent years
(see Ryabchikova et al. \cite{rya02}). Zeeman-Doppler-Imaging in all 4
Stokes parameters (Kochukhov et al. \cite{koc04}) is beginning to
provide simultaneous magnetic {\em and} abundance maps for several
elements, making it possible to assess whether there is a correlation
between the respective topologies of the surface magnetic field and of
the chemical abundances, as predicted by diffusion theory.

As far as stellar interiors are concerned, the last decade has seen radiative
acceleration calculations based on large opacity and atomic databases,
resulting in increasingly realistic modelling of the diffusion of
chemical elements. In the outer layers of the stars, the treatment of
radiative transfer proves far more difficult and time-consuming;
likewise, large-scale calculations for a fair number of chemical elements,
involving reasonably realistic spectra, the correct treatment of blends,
and accurate formal solvers require very fast, preferably parallel
computers, especially if Zeeman splitting and magnetic polarisation is
to be taken into account (Alecian \& Stift \cite{ale02}, \cite{ale04}).
Previously, Zeeman splitting was estimated to be of merely
secondary importance in the case of Si (Alecian \& Vauclair
\cite{alv81}), so only the effects of the magnetic field on the movements
of the ions across magnetic lines were being considered (see Vauclair et
al. \cite{vhp79}). Later Babel \& Michaud (\cite{bam91a}) found moderate
amplifications of the radiative accelerations (due to Zeeman splitting)
over the zero field value. Alecian \& Stift (\cite{ale04}) investigated
the magnetic amplifications of a large number of chemical elements as a
function of field strength and of field direction in detail.
Radiative accelerations were shown to be quite sensitive to blending,
magneto-optical effects, and the Zeeman patterns of the lines involved.

Since our previous study of radiative accelerations in magnetic 
fields (Alecian \& Stift \cite{ale04}) was mainly concerned with 
the amplification effect due
to Zeeman splitting, we only considered the radiative acceleration due 
to bound-bound transitions. In the present paper, we want to address 
the problem of element stratification in stellar magnetic atmospheres 
due to microscopic diffusion. Therefore, we have to compute as accurately 
as possible all the physical ingredients that determine the fluxes of
the various particles and that eventually lead to abundance 
inhomogeneities. In the following section we present detailed 
computations of radiative accelerations, including the contributions of 
bound-free transitions and the redistribution of momentum among the 
various states of ionisation of a given element. In the subsequent 
section, we consider the diffusion velocity in the presence of 
inclined magnetic fields and discuss the basic mechanism 
for the build-up of abundance stratification. Finally, we outline 
the trends of abundance stratification in the magnetic atmosphere of a 
main-sequence star with effective temperature $T_{\rm eff} = 12000$\,K.

\section{Radiative accelerations}
\label{sec:accel}

In Alecian \& Stift (\cite{ale04}), we only considered the average 
acceleration of each element due to bound-bound transitions of ions. 
Actually, to compute the diffusion velocity, one also has to consider  
the acceleration due to bound-free interactions and correct 
the average acceleration for the redistribution of momentum among the 
various ionisation states.

\subsection{Accelerations due to b-b and b-f transitions}
\label{subsec:bound}

An ion $A^{+i}$ can be radiatively accelerated both through 
bound-bound transitions -- discussed in detail in Alecian 
\& Stift (\cite{ale02}, \cite{ale04}) -- and through bound-free 
transitions (photoionisations) of the ion $A^{+(i-1)}$
(see Alecian \cite{ale94})
\begin{eqnarray}
g_{bb,i} & = & a_i \sum_{k,m>k} N_{i,k} \int_{0}^{\infty} \!
          \sigma_{i,km} \, \Phi_{\nu} \, {\rm d}\nu \label{eq:bb}\\
g_{bf,i} & = & a_i \,\sum_{k} \, N_{i-1,k} \int_{\nu_{i-1,k}}^{\infty} 
           \!(1 - y_{i-1,k}) \, \sigma_{i-1,k} \, \Phi_{\nu} \, 
           {\rm d}{\nu}\label{eq:bf}
\end{eqnarray}
where $a_i = 1 / (N_{i}\,m_A\,c)$. The mass of the atom is $m_A$, the 
respective total and k-level number densities of ion $A^{+i}$ are 
denoted by $N_{i}$ and $N_{i,k}$, and $c$ is the velocity of light. 
$\Phi_{\nu}$ is the radiative flux (in erg\,cm$^{-2}$\,s$^{-1}$\,Hz$^{-1}$),
$\sigma_{i,km}$ the cross-section of transition $k \rightarrow m$, 
and $\sigma_{i,k}$ the photoionisation cross-section with threshold
frequency $\nu_{i,k}$. The correcting factor $y_{i,k}$ has been 
introduced by Michaud (\cite{mic70}) and gives the fraction of 
momentum imparted to the electron. Usually it is approximated by 
$1.6\,(1 - \chi_{i,k} / h{\nu})$ with $\chi_{i,k}$
the ionisation energy (Sommerfeld \& Schur, \cite{som30}). Although recent
analytical studies by Massacrier (\cite{mas96}) have improved the knowledge
of $y_{i,k}$ for H-like ions, it is poorly known for the rest, and
we kept the usual
approximation. The fraction of momentum transferred to the ion is thus 
1 at threshold, goes to zero at $h{\nu} = \frac{8}{3}\,\chi_{i,k}$, and 
becomes negative at higher frequencies. There is no special treatment 
for autoionisation transitions in our approach; some of them are 
included in the photoionisation cross-sections provided by TOPbase 
(see Sect. \ref{sec:inputphys}).

The CARAT code (for a description see Alecian \& Stift \cite{ale02}, 
\cite{ale04}) 
carries out full opacity sampling of Zeeman split spectral lines 
and determines the radiative flux by solving the polarised radiative 
transfer equation. The model atmospheres used in our study have been 
calculated with the Atlas9 model atmosphere code (Kurucz \cite{kur93}), 
so the continuous opacities are based on the Atlas9 opacity 
routines. The photoionisation cross-sections $\sigma_{i,k}$, however, 
are taken mainly from TOPbase as discussed below in Sect. 
\ref{sec:inputphys}. This introduces a slight inconsistency: 
in contrast to the bound-bound case (Eq.\,(\ref{eq:bb})) where 
cross-sections, fluxes, and resulting accelerations are based on 
the same set of atomic data -- in our study we used data from
VALD (Piskunov et al. \cite{pkr95}) -- photoionisation cross-sections
$\sigma_{i,k}$ and flux $\Phi_{\nu}$ respectively in Eq.\,(\ref{eq:bf}) 
involve different atomic data sets.
Still, we consider that such an 
inconsistency has no significant consequences on the diffusion 
velocities in the present work, since Atlas9 models have proven to yield
fairly realistic stellar fluxes. Random errors may, however, stem
from uncertainties in the computed wavelengths of photoionisation
thresholds, which can be much larger than the typical Doppler widths of
spectral lines.
Pending the completion of a new stellar atmosphere code currently 
being undertaken by Stift and collaborators, we are convinced that at 
present there is no viable alternative to our approach.

\subsection{Redistribution of momentum among ions}
\label{subsec:redistrib}

In Alecian \& Stift (\cite{ale04}), we were mainly interested in 
Zeeman amplification and we approximated the acceleration of a 
given element by the simple average value of the accelerations of its 
various ions, i.e. by the sum of the individual accelerations $g_i^{rad}$ 
of the ions, weighted by their respective relative populations. This was 
justified in that particular context, since we only wanted to estimate the 
effect of magnetic fields on the accelerations. However, this kind of 
simple average is based on an implicit hypothesis, viz. it supposes that 
from the point of view of the transport process, the ions of a given element
behave like particles that are completely independent of each other.
In other words, it is assumed that they interact only by means of elastic 
collisions. Ionisation/recombination processes, however, are not elastic 
interactions, in particular because particles are not of the same type 
before and after the interaction.
What is called population $A^{+i}$ actually designates (in the 
statistical sense) the average fraction of atoms $A$ that are, at a 
given time, in the $i$th state of ionisation. From the microscopic point 
of view, in a stellar layer where ion $A^{+i}$ is dominant, the atoms $A$ 
spend most of their life in state $i$, but also to a lesser degree in the 
adjacent ionisation states $i-1$ and $i+1$. If momentum is gained by
an ion in state $i$ through absorption of a photon, and if the time 
required to lose this momentum (i.e. transfer it to the medium by means 
of collisions) is larger than the mean time for ionisation or 
recombination, a non-negligible part of this momentum will be shared 
with states $i-1$ or $i+1$. This is approximately equivalent to the
assumption that the corresponding atomic radiative force will be exerted 
mainly on state $i$ but also to a lesser degree on the ionisation 
states $i-1$ or $i+1$, which result from recombination and ionisation 
respectively.

This picture applies to the momentum gained through bound-bound 
transitions. In a bound-free transition, one assumes that the 
ionisation time is zero, so most of the momentum is acquired by ion
$A^{i+1}$ but for the momentum carried away by the ejected electron.
That {\em immediate} momentum redistribution for bound-free 
photoabsorption is implicit in Eq.(\ref{eq:bf}).

The {\em redistribution effect} was first introduced by Montmerle \& 
Michaud (\cite{mom76}, hereafter referred to as MM). It must be taken into
account in the diffusion 
process, especially when the neutral state of an element is 
involved, because the diffusion coefficient for neutrals is several 
orders of magnitude larger than for ions (in layers where hydrogen 
is strongly ionised). Actually, {\em redistribution} of momentum
raises complex physical questions that have not yet been completely solved. 
One of 
the main difficulties lies in the fact that the redistribution of momentum
must depend on the energy levels, since the probability of ionisation is higher
from higher levels than from lower ones. In the original work of MM,
this was not considered. Gonzalez et al. (\cite{gon95}, hereafter referred 
to as GLAM) studied this problem by means of a detailed study of carbon 
interaction rates (in the optically thick case) as a function of the energy 
levels. These authors also proposed an approximate method of estimating 
the redistribution of momentum. They found that the momentum of 
C{\sc ii} is mainly redistributed to C{\sc iii} when the absorption of
photons leads to final energy levels with principal quantum numbers
greater than or equal to 3, \rm
which corresponds to levels with an energy that is higher than
$2/3$ of the ionisation potential, and proposed to generalise these 
findings to all elements. Hereafter, we shall call this method the
{\em light-GLAM} method. Proffitt et al. (\cite{pro99}) applied the
detailed GLAM method to the study of Hg throughout the atmosphere. This is
certainly the most accurate calculation of the radiative acceleration 
ever done for stellar atmospheres since the detailed interaction rates
have been calculated for each energy level of Hg. In the present work 
and according to the present status of our CARAT code, we cannot apply 
such a sophisticated method. We therefore computed the radiative 
accelerations using both the {\em MM} and the {\em light-GLAM} method 
(see the discussion in Sect. \ref{sub:comprad}). 

Details of the formulae, and equations used for estimating the
respective interaction rates (or time scales)
-- collision, ionisation, and recombination --
and for computing the radiative accelerations with 
both the {\em MM} and the {\em light-GLAM} approach are presented 
in Appendixes \ref{apMM}, \ref{apGLAM}, and \ref{ap_iorecomb}.

\section{CARAT : added input physics}
\label{sec:inputphys}

The CARAT code developed by Stift (Alecian \& Stift \cite{ale02}, 
\cite{ale04}) and based on the COSSAM code of Stift (\cite{sti00}),
also discussed in Wade et al. (\cite{wbk01}), has in the past only 
been capable of calculating 
accelerations due to bound-bound transitions. Since bound-free 
processes had not been included, actual diffusion {\em velocities}
could not be determined. For the present study, we have
therefore modified and extended CARAT to include routines that
calculate photoionisation and collisional ionisation rates (see 
Appendix~\ref{ap_iorecomb}) using several different sources. Tables of 
photoionisation cross sections as a function of energy come from 
TOPbase (Cunto \& Mendoza \cite{cum92}, Cunto et al. \cite{cum93}); 
sufficient resolution of the resonances can lead to large sets of
data points for many levels. Energy level data that can be used 
in conjunction with hydrogenic formulae (see also Alecian \cite{ale94}) 
are extracted from TOPbase, from the NIST Atomic Spectra 
Database (http://physics.nist.gov/PhysRefData/ASD/index.html)
and from Kurucz (\cite{kur93}).

As pointed out above, TOPbase data for ions $A^{+i}$ with charge $i$ 
(http://vizier.u-strasbg.fr/TOPbase) consist of either detailed 
photoionisation cross section tables with
\begin{itemize}
\item $kSLP$   the quantum numbers of the spectroscopic series containing 
               the level $k$, coded as $(2S+1)*100+L*10+P$ ($P=1$
               corresponds to odd parity),
\item $kLV$    the energy position of level $k$ within its spectroscopic series,
\item $E_R$    the level energy in Rydbergs (relative to the ionisation limit),
\item $NP$     the number of subsequently listed energy -- cross-section pairs,
\end{itemize}
or of level listings which, in addition to $kSLP, kLV$ and $E_R$, include
\begin{itemize}
\item $g_{i,k}$  the statistical weight of level $k$,
\item $n_{i,k}$  the effective quantum number.
\end{itemize}
In contrast, level data from NIST or Kurucz give
\begin{itemize}
\item $J$    the total quantum number of level $k$,
\item $E_g$  the level energy (in cm$^{-1}$), measured from the ground state.
\end{itemize}
CARAT does not require any of the further atomic level data that are
provided by TOPbase, NIST or Kurucz, such as nuclear charge, number of 
electrons, electron configuration, spectroscopic term, quantum defect, 
Land{\'e} factors, and radiative lifetimes of the levels. Statistical 
weights of the levels, when not listed, are either calculated from
$kSLP$ or from $J$. 

When only energy level data are available, we resort to a hydrogenic 
approximation of the photoionisation cross section (in cm$^{-2}$) 
from level $k$
\begin{eqnarray}
\sigma_{i,k} = 7.9\,10^{-18} \frac{n_{i,k}}{(i+1)^2} 
    \left( \frac{\chi_{i,k}}{h \nu} \right)^3
\end{eqnarray}
where $i$ is the charge (the degree of ionisation) of the atom with
$i=0$ for neutrals, $\chi_{i,k}$ the ionisation potential measured
from level $k$, and $h \nu$ the energy of the photon.

CARAT offers several options for the inclusion of photoionisation 
cross section data from the various sources. Presently we always 
use the complete set of TOPbase level data. Whenever detailed cross 
section data are available, they replace the level data -- for the
purpose of identifying the levels to be replaced we need the
quantities $kSLP$ and $kLV$. The elements covered by TOPbase are 
Li, Be, B, C, N, O, Ne, Na, Mg, Al, Si, S, Ar, Ca, and Fe.
The NIST data are taken solely for those remaining elements where 
data are available for at least the first 3 ionisation stages,
viz. F, P, Cl, K, Sc, Ti, V, Cr, Mn, Co, Ni, Cu, Zn, and Ga. Fe
constitutes a special case in that TOPbase does not list the first
2 ionisation stages; they are taken from NIST. We have omitted the 
rare earths.

\section{Diffusion velocity and element stratification}
\label{sec:diffvelo}

The diffusion velocity $V_{D_{i}}$ of an ion in a binary mixture (protons,
electrons, and trace ions), without magnetic field (Aller \& Chapman 
\cite{all60}, Michaud \cite{mic70}), can be approximated by:
\begin{equation}
\label{VDi}
V_{D_{i}}\approx D_{i}\left[{-{{\partial\,\rm{ln} \,c_{i}}
\over{\partial r}}+{{A\,m_{p}}\over{kT}}\left({g{_{i}}^{\!rad}-g}\right)+
{{\left({Z_{i}+1}\right)m_ {p}}\over{2 kT}}\,g}\right],
\end{equation}
\noindent where
\begin{itemize}
\item $D_i$ is the diffusion coefficient of the ion $i$,
\item $A \, m_{\rm p}$ the mass $({\rm cgs})$ of the atom,
\item $r, T, k$ are the geometrical depth, the temperature, and
   the Boltzmann constant, respectively,
\item $c_i$ is the ion concentration (the ratio of the ion partial pressure 
   over the total gas pressure),
\item $Z_i$ the charge ($Z_i = 0$ for the neutral state)
\item $g{_{i}}^{rad}$ the {\em redistributed} radiative acceleration of the ion,
\item $g$ the gravity.
\end{itemize}
The third term in the brackets, introduced by Aller \& Chapman 
(\cite{all60}), constitutes a minor correction due to the effect 
of the electric field. In this work, we have neglected thermal 
diffusion; the first term in brackets -- ordinary diffusion due 
to a gradient in the concentration of ions --  is small (due 
to the ionisation gradient only) since in the present study
we are considering homogeneous abundances. A more sophisticated 
method of computing microscopic diffusion for a multi-component gas has 
been used in stellar evolution codes (Richer et al. \cite{ric98}). 
The approximate Eq.\,(\ref{VDi}) assumes that ions are trace particles
and is accurate enough for our purposes. The diffusion coefficients
have been discussed in detail by Paquette et al. (\cite{Paq86}) and
Michaud \& Proffitt (\cite{MichaudPr93}); here 
we use the same approximations as Landstreet et al. (\cite{lan98}) 
and Budaj \& Dworetsky (\cite{bud02}), corresponding to the adoption 
of the coefficient for binary ion-proton collisions, corrected for 
collisions with neutral hydrogen. The latter correction could prove 
significant in atmospheres with effective temperatures lower than 10000\,K.

In Eq.\,(\ref{VDi}), we have also neglected ambipolar diffusion
(Babel \& Michaud, \cite{bam91c}). This effect might be important
in places where $N$(H{\sc i})/$N$(H{\sc ii}) $\approx 1$; however, this 
does not occur in the model we will consider.

\subsection{Diffusion velocity in magnetic fields}
\label{subsec:diffvelomag}

The approximate theory of diffusion in magnetic fields can be found 
in Chapman \& Cowling (\cite{cha70}). The diffusion velocity
(\ref{VDi}) of charged particles, orthogonal to magnetic lines, is 
reduced by the factor:

\begin{equation}
f_{slow,i}={\left({1+\omega_{i}^{2}t_{i}^{2}}\right)}^{-1},
\label{fslow}
\end{equation}

\noindent
where $t_{i}$ is the collision time (the time taken by a particle $i$ 
with mass $m_i$ and charge $Z_{i}e$, to deviate by ${\pi}\over{2}$ 
through collisions from its initial motion), ${\omega_i}/{2\pi}$ the 
cyclotron frequency in a field with strength $H$:
\begin{equation}
\omega_{i}={{Z_{i}eH}\over{m_{i}c}}.
\label{omega}
\end{equation}
The collision times $t_{i}$ can be estimated easily from the diffusion
coefficients (see Appendix \ref{apMM}).

The average vertical diffusion velocity (in the presence of horizontal
magnetic fields) of an 
element can be approximated by the following expression (the sums are 
over the ionisation states $i$):

\begin{equation}
V_{D}\approx{{\sum{\;N_{i}\;f_{slow,i}\;V_{Di}}}\over{\sum{N_{i}}}}
\label{vD}
\end{equation}
\noindent
where the $N_i$ are the number densities of the respective ions.
 
This formula is the one used by Vauclair et al. (\cite{vhp79}) to study 
the Si accumulation in magnetic Ap stars. It is worthwhile to recall 
here how these authors have analysed the behaviour of the diffusion velocity 
as a function of magnetic field direction, since it is very instructive. 
Silicon is found to be overabundant in magnetic Ap stars but more or 
less normal in non-magnetic peculiar stars. Vauclair et al. (\cite{vhp79}) 
have shown that the radiative acceleration on ionised Si is not very 
strong (due to saturation of lines)
and cannot therefore lead to high overabundances, even though
the acceleration on the neutral state is strong. This explains the 
quasi-normal abundance of Si in the non-magnetic stars. But in magnetic 
Ap stars, the downward diffusion flux due to the settling of Si ions
(mainly Si{\sc ii}) is 
slowed down by the magnetic field (small $f_{slow,i}$ in Eq.\,(\ref{fslow})), 
while for the neutral state, $f_{slow,0}=1$ ensures a positive velocity 
$V_{D}$ and leads to Si overabundances. The authors concluded that the 
diffusion model is compatible  with the Si abundances observed in peculiar 
stars. Their study has been extended to the oblique rotator model by Michaud 
et al. (\cite{mic81}), assuming a dipole plus quadrupole magnetic structure. 
These authors showed how elements can accumulate, depending on the 
orientation of the magnetic field (at the magnetic poles and/or equator).

So far, our considerations have essentially been restricted to the
average diffusion velocity across horizontal magnetic lines. However, 
at the surface of magnetic stars, the direction of the magnetic field 
lines varies from 0 to ${\pi}\over{2}$  between magnetic pole(s) and 
equator(s). This angular dependence can explain inhomogeneous 
abundance distributions over the stellar surface: the sensitivity of 
the diffusion velocity to the inclination of the magnetic field lines 
determines the size of the abundance structures (see Michaud et al.
\cite{mic81}, and Alecian \& Vauclair \cite{alv81}). In magnetic fields 
that are neither strictly horizontal nor strictly vertical, the diffusion 
velocity vector is now inclined with respect to the vertical ($z$-axis). 
The inclination is given by the expressions derived by Alecian \& Vauclair 
(\cite{alv81}) (please note that their $\theta$ angle was defined with 
respect to the horizontal and not to the vertical as in the present 
work). The vertical ($z$-axis) component is:

\begin{equation}
V_{Hz,i}\approx
V_{D_{i}}\left[{f_{slow,i}+\left(1-f_{slow,i}\right)\cos^{2}\theta}\right],
\label{Vz}
\end{equation}

\noindent
and the horizontal one ($x$-axis) is:

\begin{equation}
V_{Hx,i}\approx{{1\over{2}} \, {\left(1-f_{slow,i}\right)V_{D_{i}}}}\sin2\theta.
\label{Vx}
\end{equation}

The diffusion velocity in a magnetic field is given by the vector ${\bf V}_{Hi}
=\left({V_{Hx,i},V_{Hz,i}}\right)$. For vertical magnetic lines ($\theta=0$),
one verifies that ${\bf V}_{Hi}=\left({0,V_{D_{i}}}\right)$ according
to Eqs.\,(\ref{Vz}) and (\ref{Vx}). In this case, the only effect of the
magnetic field comes from the amplification of the vertical component of
radiative acceleration due to Zeeman splitting (Alecian \& Stift \cite{ale04}). 
For horizontal magnetic field lines ($\theta={{\pi}\over{2}}$), one finds  
${\bf V}_{Hi}=\left({0,f_{slow,i}V_{D_{i}}}\right)$ as in Eq.\,(\ref{vD}). For 
inclined magnetic lines, the horizontal component given by expression (\ref{Vx}) 
represents only a horizontal diffusion of ions (keep in mind that the horizontal 
component of the radiative acceleration is generally negligible, see Alecian \& 
Stift \cite{ale04}). However, this horizontal component of the diffusion 
velocity, which could be of the same order of magnitude as the vertical one, 
has a very small effect on the horizontal distribution 
of elements. This results from a combination of time scales and of spatial
scales. Indeed, in the case of silicon, for instance, Megessier (\cite{meg84})
has estimated that Si migration from the pole to the equator would take some 
$10^7$ years! Only if the life time of the magnetic structures is of this 
order of magnitude, can an observable effect of horizontal diffusion
be produced. Because of the much smaller vertical height scales, vertical 
diffusion leads to observable effects much earlier (by 4 to 5 orders
of magnitude). One can thus conclude that it appears rather unlikely
that horizontal 
diffusion will have a significant effect. This means that the abundance 
inhomogeneities on magnetic Ap stars (patches or rings) are mainly due
to the angular dependence of the vertical component of the velocity
vector as given by Eq.\,(\ref{Vz}), combined with some local mass-loss 
velocity. Therefore we shall stick hereafter to the one-dimensional 
scalar form of Eq.\,(\ref{vD}) and just use $V_{H,zi}$ instead of 
$f_{slow,i} \, V_{Di}$. All the computations assume zero mass-loss velocity.

\subsection{Abundance stratification}

One of the main goals of Ap star modelling is the simulation of the
development in time of abundance stratifications.  Unfortunately, 
this is still beyond our possibilities, the stratification process 
being non-linear and requiring both further theoretical (and
software) developments and huge computing power. Still, a 
first guess of the kind of stratifications to be expected can be 
obtained by considering the flux of the various diffusing elements
at the beginning of the process ($t=0$), when elements still have their 
normal abundances and with no competing processes considered. We
discuss this point in more detail in Sect. \ref{sec:discuss}. 
Such a study was first carried out by Alecian \& Vauclair 
(\cite{alv81}) who neglected the Zeeman effect. In the present work, 
by neglecting the horizontal flux as discussed in the previous subsection,
we compute the vertical diffusion flux $F_D$, at depth $z$ (increasing 
towards the centre of the star) by
\begin{equation}
F_{D}(z)=\sum{N_{i}\,V_{Hz,i}},
\label{FD}
\end{equation}
\noindent
from which we determine, for each element considered, the local 
rate of abundance change per unit time, at time $t=0$. This rate is
$a_0  = {{\partial _t N} \mathord{\left/
 {\vphantom {{\partial _t N} N}} \right.
 \kern-\nulldelimiterspace} N}$, 
where $N$ is the number density of element $A$, i.e. $\sum N_i$.
Let $N_{\rm slab}$ be the total number of particles $A$ in a cylinder of height
$\Delta z$ and section unity. One has:

\begin{equation}
\frac{{\partial _t N}}{N} \approx \frac{{\Delta _t N_{slab} }}{{N_{slab} }} =
\frac{{F_D \left( {z + \Delta z} \right) - F_D \left( z \right)}}{{N\Delta z}},
\label{dNvsN}
\end{equation}

\noindent
which leads to
\begin{equation}
a_0=N^{-1}\,\partial_{z}\,F_{D}(z).
\label{at}
\end{equation}

\section{Results and discussion}
\label{sec:discuss}

We have computed the radiative accelerations and diffusion velocities
of 28 elements (Be to Ga) that are generally observed in Ap stars, and 
we have enough atomic data at our disposal for most of these 
to obtain reasonably accurate results. The main-sequence stellar model we 
adopted comes from Kurucz (\cite{kur93}) with $T_{\rm eff} = 12000\,K$, $\log
g = 4.00$, and solar abundances.

\subsection{Computation of the radiative accelerations}
\label{sub:comprad}

Hereafter, we consider the total radiative acceleration on each element,
computed by both the {\em MM} and the {\em light-GLAM} methods, with

\begin{equation}
g_{\rm tot}^{rad}  = \frac{{\sum {N_i \, D_{Hz,i} \, g_i^{rad} } }}
{{\sum {N_i \, D_{Hz,i} } }}
\label{eq:gtot}
\end{equation}

\noindent
where
\begin{equation}
D_{Hz,i}  = D_i \left[ {f_{slow,i}  + \left( {1 - f_{slow,i} }
\right)\cos ^2 \theta } \right].
\label{eq:DHz}
\end{equation}

When the total radiative acceleration (vertical component) equals 
gravity and if concentration gradients are
neglected, the diffusion velocity of the element should become zero
(according to 
Eqs.\,(\ref{VDi}) and (\ref{FD})). This leads quite
naturally to the definition of an effective total acceleration
$g_{\rm tot}^{eff}$ (taking into account the correction due to the 
electric field):

\begin{equation}
g_{\rm tot}^{eff}  = \frac{{\sum {N_i \, D_{Hz,i} \left[ {g_{i}^{rad} 
- \left( {1 - \frac{{Z_i + 1}}{{2A}}} \right)g} \right]} }}
{{\sum {N_i \, D_{Hz,i} } }}.
\label{eq:geff}
\end{equation}
\noindent
Note that the $g_i^{rad}$ are the respective redistributed accelerations. 
The total diffusion flux (\ref{FD}) can be expressed directly as 
a function of $g_{\rm tot}^{eff}$:
\begin{equation}
F_D \left( z \right) \approx N\left\langle D \right\rangle 
\frac{{A\,m_p }}{{kT}}\,{g_{\rm tot}^{eff}},
\label{eq:Feff}
\end{equation}

\noindent
where we have defined the average diffusion coefficient as:
\begin{equation}
\left\langle D \right\rangle  = \sum {N_i \, D_{Hz,i} \,/\,N}.
\label{eq:Dmoy}
\end{equation}
\begin{figure}
\includegraphics[height=6cm]{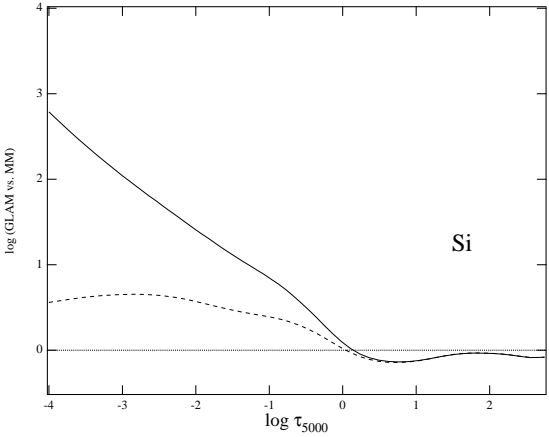}
\caption{Comparison of the Si accelerations obtained by the {\em light-GLAM}
method and the {\em MM} method (Eq.\,(\ref{eq:gtot})). The logarithm of
$g_{GLAM}^{rad}/{g_{MM}^{rad}}$ is shown vs. the logarithm of optical
depth for a model with $T_{\rm eff} = 12000\,K$, $\log g = 4.00$.
The dashed line pertains to 0 Gauss, the solid line to 10000 Gauss.
}
\label{fig:4558fig1}
\end{figure}

According to the study by GLAM, one expects to have large discrepancies 
between the {\em MM} and the {\em light-GLAM} methods, and we confirm this
trend. In Fig.\,\ref{fig:4558fig1} we present the logarithm 
of the ratio of the accelerations {\em light-GLAM} vs. {\em MM} for silicon
(to be discussed in some detail in Sect. \ref{sub:SiCase}). 
The discrepancy is especially strong in the magnetic case and for
layers with optical depth smaller than unity. We carried out such a
comparison for all the 28 elements investigated, and found that the
discrepancy is much smaller for some of them, 
but larger for some others. The elements for which
the discrepancies are smaller than 0.3\,dex in the non-magnetic case are: 
Be, B, Ca, Sc, Ti, V, Cr, Mn, Fe, Co, Ni, Cu. The discrepancies are generally
larger in the magnetic case. We do not notice obvious systematic trends except
that {\em MM} is larger than {\em light-GLAM} for noble gases (Ne, Ar) and
that {\em MM} is smaller than {\em light-GLAM} for the alkali elements
(Na, K), {\em i.e.} elements with a noble gas configuration in their 
first ionisation stage. This reveals the role of the distribution of energy 
levels, since levels are not considered in the same way in {\em light-GLAM} 
and in {\em MM}. Silicon appears to be a rather typical case concerning the 
discrepancy between the {\em MM} and the {\em light-GLAM} methods; we
discuss it in more detail. Figure\,\ref{fig:4558fig1} shows that in the 
non-magnetic case, the {\em MM} acceleration of Si is smaller 
by 0.66\,dex, which is the same order of magnitude as the discrepancy 
found by GLAM for carbon -- their discrepancy, however, goes in the opposite
direction, their {\em MM} acceleration being stronger. Notice that we cannot 
compare our results concerning carbon with those of GLAM since their
computations apply to a star with $T_{\rm eff} = 8000\,K$ and to deep 
layers (outside the validity domain of CARAT). To better understand the 
reasons for such large differences, we examined the 
contributions 
of the various terms entering both methods in detail, especially for the neutral
state since it plays a major role in momentum redistribution as already
mentioned in Sect. \ref{subsec:redistrib}. The {\em light-GLAM}
method assumes that ionisation is dominated by transitions to the continuum
from high energy levels, because after absorption of a photon leading 
to a sufficiently high-lying energy level, collisional excitations prevent
de-excitation. For the {\em MM} method, we have considered the total
direct ionisation rate from any level (see Eq.(\ref{betaion})), without
taking the added path of excitation collisions before
ionisation into account. 
The {\em MM} method thus gives too much weight to low energy
levels. If one considers that the {\em MM} method underestimates the
redistribution of momentum from the neutral to the first ionisation state,
one is led to conclude that it gives an upper limit of the radiative
acceleration. On the other hand, the {\em light-GLAM} method is certainly
less justified in the upper atmosphere, where collisional excitations
decrease with density, while photoionisation remains efficient. We
checked this last argument by comparing the average ionisation rates
from low levels (below $2/3$ of the ionisation potential) with those from
higher levels, but the final word will come from the detailed analysis
of ionisation times, level by level, and using a detailed method like
the one of Proffitt et al. (\cite{pro99}), which will be done in future
work.

Clearly, we are confronted with a serious dilemma: which method to choose
when they yield such different results without a definitive argument in
favour of one or the other? It seems reasonable to consider that the 
{\em MM} method gives an upper limit of the radiative acceleration; 
however, when it becomes larger than the acceleration given by the 
{\em light-GLAM} method, the correct acceleration should lie somewhere 
in between. To compute the diffusion velocities used in the next 
subsections, we thus decided to use $g_i^{rad}$ values given
by the minimum of the respective {\em MM} acceleration and of the 
geometrical mean of the acceleration obtained by both methods:

\begin{equation}
g_i^{rad}  = {\rm{Min}}\left( {g_{MM,i}^{rad} ,\sqrt {g_{MM,i}^{rad}  \cdot
g_{GLAM,i}^{rad} } \;} \right)
\label{eq:gtotav}
\end{equation}

\noindent
where $g_{MM,i}^{rad}$ and $g_{GLAM,i}^{rad}$ stand for the radiative 
accelerations obtained by the respective methods.

\begin{figure*}
\centering
\includegraphics[width=17cm]{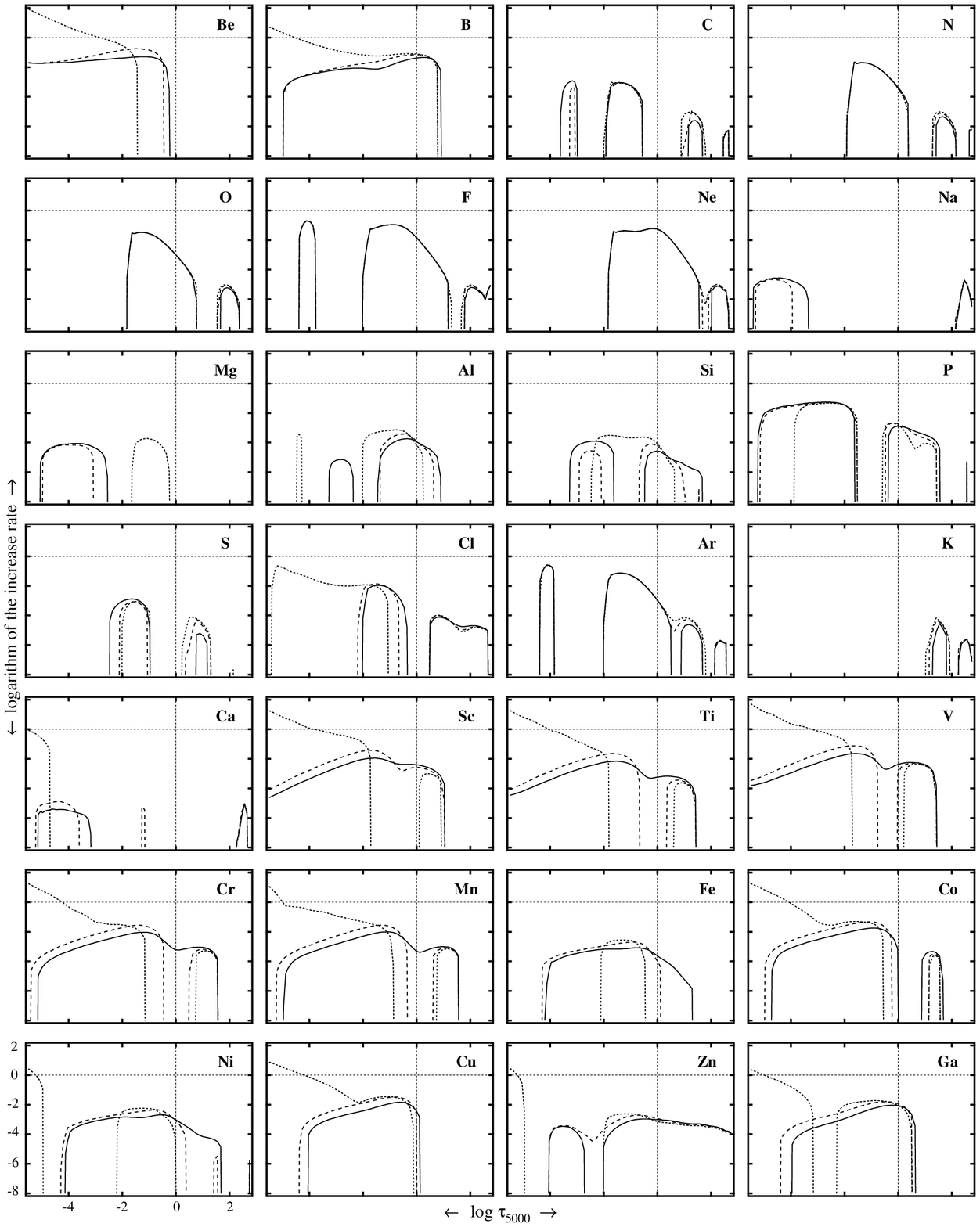}
\caption{
%Same as Fig.\ref{fig:Layout_abrate0d} for a horizontal magnetic field
%($90\,\degr$). 
Logarithm of the rates of abundance increase (per year) vs. logarithm of
the optical
depth (at 5000\,{\AA}) for a horizontal magnetic field ($90\,\degr$). All plots
have the same scale as the one in the bottom left corner. Solid lines pertain to
$20000\,G$, dashed lines to $10000\,G$, and dotted lines to $0\,G$.
}
\label{fig:4558fig2}
\end{figure*}
\begin{figure*}
\centering
\includegraphics[width=17cm]{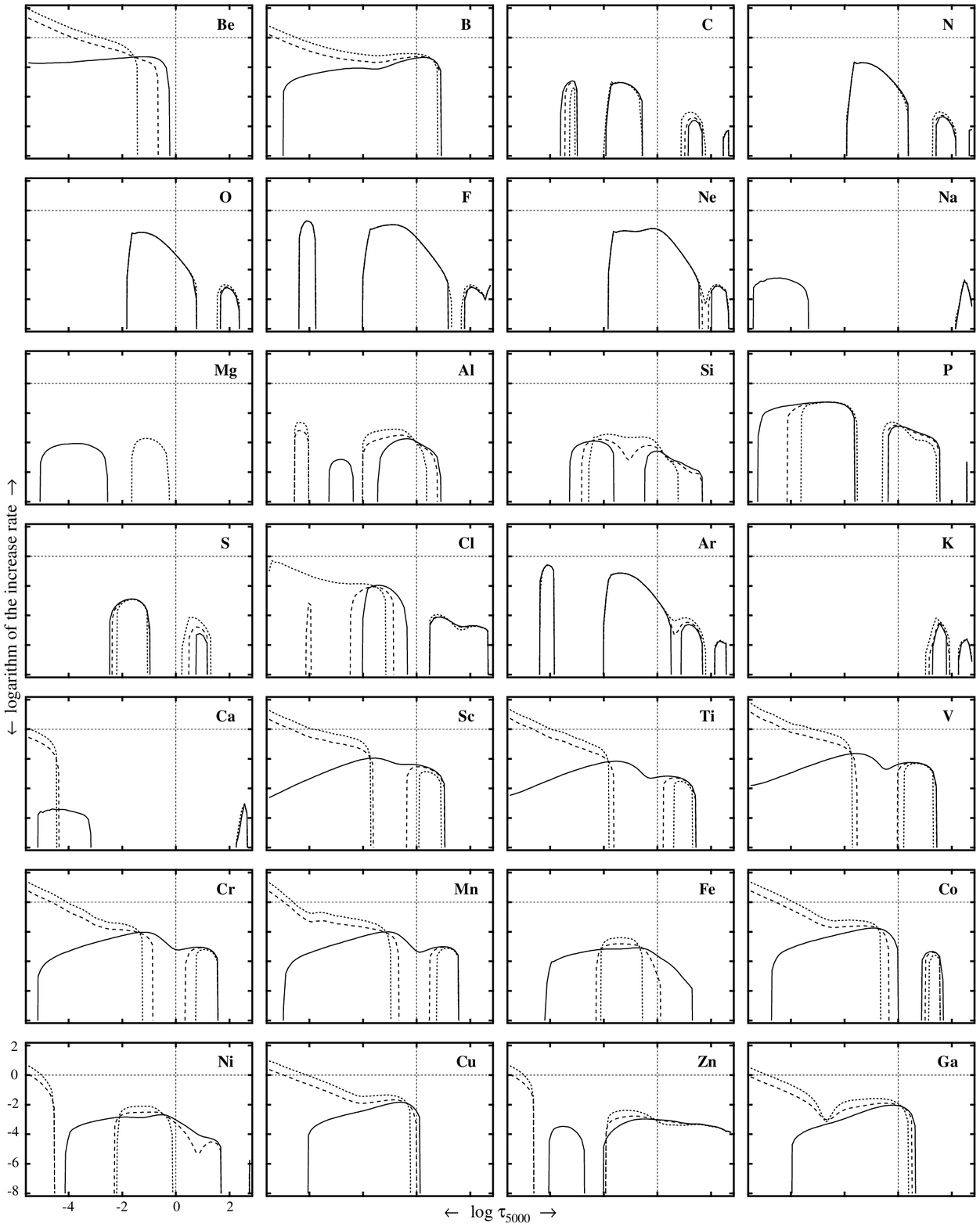}
\caption{
Same as Fig.\,\ref{fig:4558fig2} but for a magnetic field of $20000\,G$
with various inclinations, viz. $90\,\degr$ (solid lines), $60\,\degr$ (dashed
lines), and $0\,\degr$ (dotted lines).
}
\label{fig:4558fig3}
\end{figure*}

\subsection{Estimates of stratification trends}
\label{sub:strattrends}

The respective rates of abundance increase for each element are presented 
in Figs.\,\ref{fig:4558fig2} and \ref{fig:4558fig3}. The 
curves show the logarithm of $a_0 (cgs) \times 3.1556 \cdot 10^7$
(secs) which is the abundance change in one year ($a_0$ is given by 
Eq.\,(\ref{at})). We have plotted this quantity only for positive values 
of $a_0$. Interpreting the curves physically, one could say that they 
give the amount by which the abundance of the element considered is 
increased in one year of continuous diffusion. Implicitly we assume that
the accelerations remain constant during this time since we use $a_0$
for a linear extrapolation of the abundance changes by simply multiplying
with $\Delta t$. This is of course a rough approximation; it is known that 
the accelerations decrease non-linearly with 
increasing abundances when lines are saturated. Furthermore, we have not
considered any
boundary conditions here as has to be done when solving the continuity
equation. Therefore the curves presented here merely display initial 
trends in abundance increases. In other words, when the plotted rate 
is equal to unity at a given depth, this means that the abundance of 
element $A$ is doubled at that depth in one year of diffusion. One could also 
consider Figs.\,\ref{fig:4558fig2} and \ref{fig:4558fig3} as 
giving an approximate picture of the accumulation of the various elements in 
the atmosphere after one year of diffusion.

Note that if the flux derivative is positive, the trend in abundance changes
can be
positive even in places where elements are not supported. This corresponds to
the accumulation of falling particles. Possibly, that kind of accumulation will
not survive in any fully time-dependent calculation.

The \emph{holes} correspond to places where the abundance decreases. Assuming 
that the abundance stratification process tends to a stationary solution
($a_\infty   \to 0$), which still has to be demonstrated, a much better 
prediction of element accumulations can be obtained through an iterative 
procedure where the accelerations are recomputed after each iteration with 
updated abundance values (see Hui Bon Hoa et al. \cite{hbh02}); this is,
however, outside the scope of the present work.

Figure\,\ref{fig:4558fig2} shows the rates of abundance changes for 
2 different values of a horizontal magnetic field (10000\,G and 20000\,G), 
as well as the zero field case (angle-independent). For some elements, the
rates of abundance changes seem to be completely insensitive to the magnetic
field. 
This is due to the fact that these elements, according to our atomic data, 
experience radiative accelerations mainly through b-b absorptions in the 
neutral state. Therefore the diffusion velocity is dominated by the neutral 
term in Eq.\,\ref{vD} for which $f_{slow,0} = 1$ holds exactly. This is, so 
far, the only systematic behaviour we have been able to clearly identify. 
For other elements, trends in accumulations are highly sensitive to magnetic
field strength and orientation through $f_{slow,i}$ and $\theta$ in 
Eqs.\,(\ref{vD}) and (\ref{Vz}). As shown in Fig.\,\ref{fig:4558fig3}, 
for a given magnetic field of 20000\,G, many elements exhibit significantly
different accumulations at the magnetic poles and equators: 
the thickness of the respective accumulation zones (or element clouds) often 
appears to be more uniform and larger near the equators than near the poles.
Keep in mind that the differences between the dotted lines of 
Fig.\,\ref{fig:4558fig2} and
Fig.\,\ref{fig:4558fig3} can only
be due to Zeeman amplification. We note that $a_0$, which depends on the 
derivative of the diffusion fluxes, is not very sensitive to the effect 
of Zeeman amplification on the accelerations. 
This is because Zeeman amplifications are more efficient in changing 
the maximum overabundance of those elements that can be supported by 
the radiation field than in changing the spatial derivatives of the 
accelerations (in Eq.(\ref{at})).
Although we expect the final amount of accumulations for 
each element to depend much more significantly on Zeeman amplifications, 
we cannot demonstrate this effect at present since, as mentioned above, 
we extrapolate the $a_0$ in a linear way.

\subsection{The silicon case}
\label{sub:SiCase}

At this point it is tempting to examine what happens to silicon and 
to compare these first results to the peculiar behaviour of Si
discussed in Sect. \ref{subsec:diffvelomag}. According to 
Figs.\,\ref{fig:4558fig2} and \ref{fig:4558fig3}, 
a cloud of Si starts to form  above 
$\tau \approx 10^{-2}$ in the magnetic case; in the non-magnetic case, Si
starts to accumulate between $\tau \approx 10^{-2}$ and $\tau \approx 10$.
This is apparently at variance with Vauclair et al. (\cite{vhp79}), who 
predict that Si does not accumulate in the zero field case (or in
a vertical magnetic field) but will accumulate only when a horizontal 
magnetic field is present.

\begin{figure*}
%\centering
\includegraphics[width=16cm]{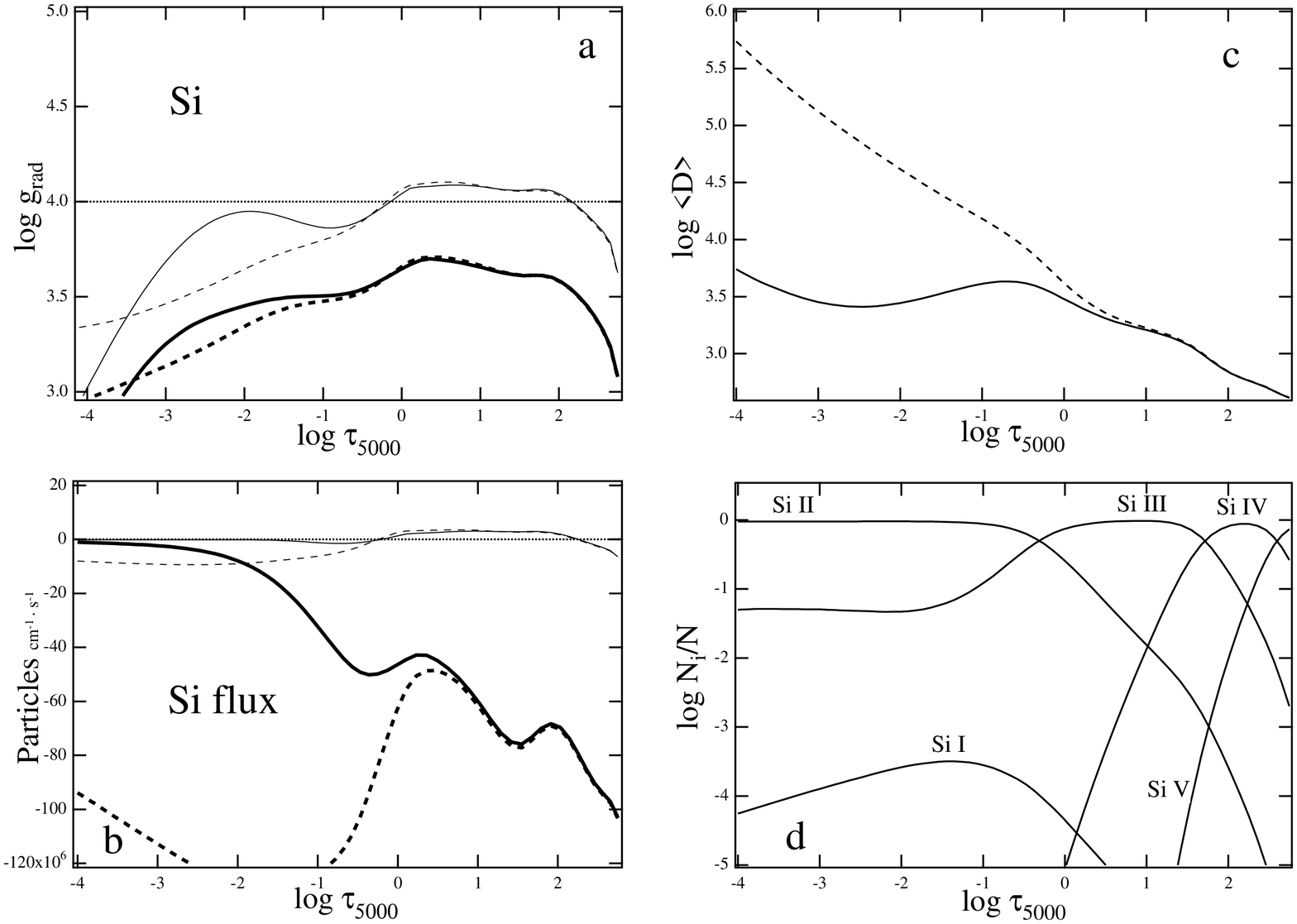}
\caption{
Detailed results for silicon. The dashed lines pertain to 0\,G, the
solid lines to 10000\,G (horizontal). The thin lines are for solar 
silicon abundance, the heavy lines for a 1\,dex overabundance 
(panels (a) and (b)).
Panel (a): Radiative acceleration of Si (Eq.\,(\ref{eq:gtotav})).
Panel (b): Diffusion flux of Si (Eq.\,(\ref{FD})).
Panel (c): Average diffusion coefficient
$\left\langle D \right\rangle  = \sum {N_i f_{slow,i} D_i \,/\,N}$ for
horizontal magnetic fields ($\cos \theta=0$).
Panel (d): Relative populations of the first five Si ions.
}
\label{fig:4558fig4}
\end{figure*}

To understand this unexpected result we 
carried out some additional computations assuming 10 times more
silicon than in the solar case (adopting 7.55 for the logarithmic
Si solar abundance on a scale where the hydrogen abundance is 12).
Figure\,\ref{fig:4558fig4} displays the relevant data for the
analysis of the behaviour of Si. \rm Panel\,(a) shows the radiative
accelerations, according to
Eqs.\,(\ref{eq:gtot}) and (\ref{eq:gtotav}). 
Our radiative accelerations are slightly stronger around $\tau = 1$ than 
those of Vauclair et al. (\cite{vhp79} (their Fig.2c)  for 0\,G, and closer to 
those of Alecian \& Vauclair \cite{alv81} (their Fig.1a). These 
differences are not surprising since our atomic data and our radiative
transfer calculations are much more accurate. As predicted, due to the 
presence of a horizontal magnetic field, more Si can be supported
than without a magnetic field for $\tau \approx 10^{-2}$, but the
radiative accelerations remain smaller than gravity for $\tau< 1$.
Silicon enhanced 10 times is not at all supported by the radiation field
throughout the atmosphere at 0\,G; despite somewhat higher acceleration 
values above $\tau \approx 10^{-1}$, this also remains true for the 
10\,kG case.
Considering the diffusion flux (according to 
Eq.\,(\ref{FD})) of silicon in panel\,(b), one can readily verify that
the fluxes are positive in places where the radiative accelerations 
exceed gravity ($\log g=4.0$). 
For the magnetic case, diffusion fluxes go to zero at small optical depths.
This can be explained by the strong decrease, above 
$\tau \approx 1$, of the average diffusion coefficient 
$\left\langle D \right\rangle$
in the magnetic case with $\cos \theta=0$, as shown in panel\,(c) --
the relative populations of the Si ions are displayed in panel\,(d). The
presence of neutrals mitigates this decrease, but not enough to
prevent the rather strong decrease in $\left\langle D \right\rangle$
due to ions.

One can conclude that Si is better supported above $\tau \approx 0.1$
by the radiation field in places where the magnetic field lines are 
horizontal, but less clearly than was
predicted by Vauclair et al. (\cite{vhp79}). The total
radiative acceleration is less enhanced by the magnetic field than
expected, and our results reveal that
Si could also be slightly overabundant in places where the magnetic field is
vertical (or zero).
On the other hand, we have to point out the limits of 
the ``{\em $a_0 \times \Delta t$}" linear analysis carried out
in this paper, since the complex Si behaviour 
is not revealed in Figs.\,\ref{fig:4558fig2} and
\ref{fig:4558fig3}.

\section{Conclusions}
\label{seq:conclusion}

For this study we have significantly increased the amount of physical 
processes included in the CARAT code.  We added the radiative 
accelerations due to bound-free transitions by using detailed b-f 
transition cross sections (whenever available), we added the 
computation of detailed ionisation/recombination rates, and took
the momentum redistribution among ions into account. This has 
allowed us to obtain results of higher astrophysical relevance than 
just radiative accelerations. We are now able to calculate diffusion 
velocities (and particle fluxes) in the presence of magnetic fields 
at arbitrary angles. Finally we estimated and discussed abundance 
stratification trends in magnetic Ap stars.

After our previous study of the magnetic amplification of radiative
accelerations, this work constitutes a second step towards the modelling
of abundance stratifications in CP stars. Several further steps are
still necessary to reach that final goal, the most critical of them being
an improvement in the theory of momentum redistribution among ions. 
The use of Eq.\,(\ref{eq:gtotav}) is unsatisfactory, at least for 
elements for which large discrepancies are found between the {\em light-GLAM}
and the {\em MM} methods (see Sect. \ref{sub:comprad}). 
We thus plan to apply a detailed GLAM method following  Proffitt et al. 
(\cite{pro99}) but adapted to the magnetic case, which will become
possible thanks to a forthcoming NLTE version of CARAT. In order to study tepid
chemically peculiar stars where neutral hydrogen 
cannot be neglected, we plan to implement ambipolar diffusion in CARAT.
A good estimate of final abundance stratifications requires more than simple
linear extrapolation of diffusion fluxes, since radiative accelerations depend
non-linearly on chemical abundances and on the element fluxes at the boundaries
(ingoing and outgoing particles at the top/bottom of the atmosphere).
Assuming that
a unique stationary solution exists in static atmospheres, an iterative method
will converge towards that solution, but in the absence of a unique and/or
stationary solution, only the simultaneous time-dependent solution of the
continuity equation will allow us to find the correct solution.

\begin{acknowledgements}
We would like to thank Georges Michaud for very useful discussions,
particularly about problems concerning redistribution of momentum,
and for a careful reading of the manuscript.
MJS acknowledges support by the {\sf\em Austrian Science Fund 
(FWF)}, project P16003-N05 ``Radiation driven diffusion in magnetic 
stellar atmospheres'' and through a Visiting Professorship at the 
Observatoire de Paris-Meudon and Universit{\'e} Paris 7 
(LUTH). Thanks go to AdaCore for generously providing us with
the GNAT\,Pro Ada95 compiler and toolsuite. A sizeable fraction of 
the calculations were carried out on the Sgi Origin\,3800 of 
the CINES in Montpellier.
\end{acknowledgements}

{}

\appendix{}

\section{The redistribution of momentum, the ``MM'' method}
\label{apMM}

The collision time for ions can be  defined as the time required for 
a deflection by $\pi /2$ through collisions with the other particles 
in the medium (mainly protons). This collision time is related to 
the diffusion coefficient (Spitzer \cite{spi68}) by:
\begin{equation}
\label{eq:t_col}
t_{i}  = \frac{{A_i \, m_p }}{{kT}}\;D_i ,
\end{equation}

\noindent
where $D_i$ is the diffusion coefficient of the ion $i$ (presented 
in more detail in Sect. \ref{sec:diffvelo}),
$A_i \, m_{\rm p}$ the mass $({\rm cgs})$ of the ion, and
$T, k$ are the temperature and the Boltzmann constant, respectively.

Let $\beta_{col,i}$ be the collision rate defined as $t_{i}^{-1}$.  
Let $\beta_{ion,i}$ and $\beta_{rec,i}$ be the respective ionisation
and recombination rates (see Appendix \ref{ap_iorecomb}). The total 
interaction rate $\beta_{tot,i}$ is then defined as 
$\beta_{col,i}+\beta_{ion,i}+\beta_{rec,i}$.

According to the redistribution model proposed by MM, and limiting the
redistribution of momentum to only the adjacent ionisation stages 
(we used the
expressions proposed by Alecian \& Vauclair \cite{ale83}), the average 
redistributed radiative acceleration of $A^{0}$ and $A^{+i}$ (for $i>0$)
may be written as:

\begin{equation}
\label{eq:grMM0}
g_0^{rad}  \approx g_{bb,0} \,\frac{{\beta _{col,0} }}{{\beta _{tot,0} }} +
g_{A^{+1}} \,\frac{{N_1 }}{{N_0 }}\frac{{\beta _{rec,1} }}{{\beta _{tot,1} }}
\frac{{\beta _{col,0} }}{{\beta _{tot,0} }},
\end{equation}

\begin{eqnarray}
\label{eq:grMM}
\nonumber g_i^{rad} & \approx &
g_{A^{+(i-1)}}{{N_{i-1}}\,\over{N_{i}}}{{\beta_{ion,i-1}}
\over{\beta_{tot,i-1}}}{{\beta_{col,i}}\over{\beta_{tot,i}}} 
 + g_{A^{+i}}\,{{\beta_{col,i}}\over{\beta_{tot,i}}} \\
& &  + \, g_{A^{+(i+1)}}{{N_{i+1}}\over{N_{i}}}{{\beta_{rec,i+1}}
    \over{\beta_{tot,i+1}}}{{\beta_{col,i}}\over{\beta_{tot,i}}},
\end{eqnarray}
\noindent where $g_{A^{+i}}=g_{bb,i}+g_{bf,i}$ (see Sect. \ref{subsec:bound}).

\section{The redistribution of momentum, the ``light-GLAM'' method}
\label{apGLAM}

In this method discussed in Sect. \ref{subsec:redistrib}, one defines 
an {\em energy segregation threshold} such that all the momentum gained 
through a b-b transition of $A^{+i}$ is kept by the ion when the
energy of the higher level is lower than a given value (here $2/3$
of the ionisation potential). We denote the corresponding acceleration by
${g'}_{bb,i}$. The momentum gained through absorption from higher energy 
levels is entirely given to the ion $A^{+(i+1)}$, and the corresponding 
acceleration is denoted by ${g''}_{bb,i}$. The average 
redistributed radiative acceleration of $A^{0}$ and $A^{+i}$ (for $i>0$) 
using the {\em light-GLAM} method 
may then be written as:

\begin{equation}
\label{eq:grGLAM0}
g_0^{rad}  \approx g'_{bb,0} ,
\end{equation}

\begin{equation}
\label{eq:grGLAM}
g_i^{rad}\approx{{N_{i-1}}\over{N_{i}}}\,{g''}_{bb,i-1}+{g'}_{bb,i}+g_{bf,i} .
\end{equation}

\section{The ionisation and recombination rates}
\label{ap_iorecomb}

The direct photoionisation rate for ion $A^{+i}$, in level $k$,  per atom is
(Mihalas \cite{mih78}):
\begin{eqnarray}
R_{ik} = 4\,\pi \int_{\nu_0}^{\infty} \frac{\alpha_{\nu} J_{\nu}\,d{\nu}}
{h {\nu}}
\end{eqnarray}
where $\alpha_{\nu}$ denotes the photoionisation cross section,
$J_{\nu}$ the mean intensity of radiation, $h$ the Planck constant,
and ${\nu_0}$ the frequency of the photoionisation edge. In LTE,
$J_{\nu}$ corresponds to the Planck function $B_{\nu}$.

The radiative recombination rate per atom can be derived from detailed 
balancing arguments and we write
\begin{eqnarray}
R_{ki} = 4\,\pi
  \int_{\nu_0}^{\infty} \frac{\alpha_{\nu}}{h {\nu}}
   \left( 2 h {\nu}^3 / c^2 + J_{\nu} \right) e^{-h {\nu} / k T} 
   d{\nu} .
\end{eqnarray}
Approximate collisional ionisation rates based on photoionisation
cross sections are given by
\begin{eqnarray}
C_{ik} = 1.55\,10^{13} \, N_e \, T^{-1/2} \, g_i \, 
   \alpha ({\nu_0}) \, \exp (-u_0) / u_0
\end{eqnarray}
where $N_e$ is the electron density, $\alpha ({\nu_0})$ the 
threshold photoionisation cross section, and 
$u_0 = h {\nu_0} / k T$ (see Mihalas \cite{mih78}). 
For the parameter $g_i$, we follow Seaton (2005, private 
communication) and assume $g_0=0.07, g_i=0.25$ (for $i>0$). Again, detailed 
balancing arguments can be applied to obtain the downward rate as
\begin{eqnarray}
C_{ki} = (N_i / N_j)^* C_{ik},
\end{eqnarray}
where the asterisk denotes LTE values.

The ionisation rate of Appendix \ref{apMM} is then given by:

\begin{equation}
\label{betaion}
\beta _{ion,i}  = \frac{1}{{N_i }}\sum\limits_k {N_{ik} 
\left( {R_{ik}  + C_{ik} } \right)} 
\end{equation}
 
The recombination rate has been evaluated assuming LTE as in Hui-Bon-Hoa
et al. (\cite{hui96}).

\end{document}